\documentclass{article}
\usepackage[utf8]{inputenc}
\usepackage{graphicx}
\usepackage{color}
\usepackage{float}
\usepackage{slashed}
\usepackage{amsthm,amsmath,amssymb,mathrsfs}
\usepackage{setspace}
\pdfoutput=1 %avoid JHEP/JCAP pdflatex error
\allowdisplaybreaks
\usepackage{jheppub}
\usepackage{xcolor} 
\usepackage{babel}
\begin{document}
%\title{On the validity of neutrino forces: from $1/r^{5}$ to $1/r^2$ and $1/r$}
%\title{On the short-range behavior of neutrino forces: \\
%$1/r^{5}$,  $1/r^{4}$,  $1/r^2$, and  $1/r$}
\title{On the short-range behavior of neutrino forces beyond the Standard Model: from $1/r^5$ to $1/r^4$, $1/r^2$, and $1/r$}
\author[a,b]{Xun-Jie Xu} 
\emailAdd{xuxj@ihep.ac.cn}
\author{and} 
\author[a,b]{Bingrong Yu} 
\emailAdd{yubr@ihep.ac.cn}
\affiliation[a]{Institute of High Energy Physics, Chinese Academy of Sciences, Beijing 100049, China}
\affiliation[b]{School of Physical Sciences, University of Chinese Academy of Sciences, Beijing 100049, China}

% \abstract{It is well-known that the long-range forces resulting from the exchange of a pair of neutrinos between two objects separated $r$ will lead to an effective potential decreasing as $1/r_{}^5$ for the contact interaction. In this paper, we embed the contact vertex into a full renormalized theory and investigate the short-range behavior of the neutrino forces. We consider two possible scenarios to open up the contact vertex by introducing a $t$-channel or an $s$-channel mediator. We derive a general formula that is valid to describe the potential in all regions as long as the external particles remain non-relativistic. In both scenarios, the potential decreases as $1/r_{}^5$ in the long-range limit as expected. In the short-range limit, the $t$-channel potential exhibits $1/r$ behavior, while the $s$-channel potential exhibits $1/r_{}^4$ and $1/r_{}^2$ behavior.    
% }
%leads to a special type of long-range forces, with the effective potentials decreasing as $1/r_{}^5$, 
\abstract{The exchange of a pair of neutrinos between two objects, seperated by a distance $r$, 
leads to a long-range effective potential proportional to $1/r_{}^5$,
assuming massless neutrinos and four-fermion contact interactions. In this paper, we investigate how this known form of neutrino-mediated potentials might be altered if the distance $r$ is sufficiently short, corresponding to a sufficiently large momentum transfer which could invalidate the contact interactions. 
We consider two possible scenarios to open up the contact interactions by introducing a $t$-channel or an $s$-channel mediator. We derive a general formula that is valid to describe the potential in all regimes as long as the external particles remain non-relativistic. In both scenarios, the potential decreases as $1/r_{}^5$ in the long-range limit as expected. In the short-range limit, the $t$-channel potential exhibits the Coulomb-like behavior (i.e. proportional to $1/r$), while the $s$-channel potential exhibits $1/r_{}^4$ and $1/r_{}^2$ behaviors.    
}
% \newpage

\maketitle

\section{Introduction}
The long-range forces arising from the exchange of a pair of neutrinos
 between two objects (the so-called neutrino forces) have been an old and interesting topic dating back to early considerations in the 1930s~\cite{manypapers}, subsequently followed by quantitative calculations in the 1960s~\cite{Feinberg:1968zz}.
In the framework of the Standard Model (SM), the (spin-independent part of the) effective potential between two electrons, induced by massless neutrino exchange, is formulated as~\cite{Feinberg:1968zz,Feinberg:1989ps, Hsu:1992tg}:
\begin{eqnarray}
\label{eq:SM nu-force}
V_{ee}^{}(r)=\left(2\sin_{}^2 \theta_{\rm W}^{}+\frac{1}{2}\right)_{}^2 \frac{G_{\rm F}^2}{4\pi^3}\frac{1}{r^5}\;, 
\end{eqnarray}
where $G_{\rm F}^{}$ is the Fermi constant,
$\theta_{\rm W}^{}$ is the Weinberg angle, and $r$ is the distance between two electrons. Early derivations~\cite{Feinberg:1968zz,Feinberg1965, Feinberg:1989ps} by Feinberg and Sucher employed the dispersion technique\footnote{See also Ref.~\cite{Segarra:2015mqp} for a more pedagogical introduction.} and the result was verified by other authors via different approaches such as the Fourier transform~\cite{Hsu:1992tg} and the Hamiltonian formalism~\cite{LeThien:2019lxh}.

As has been well established by neutrino oscillation experiments, neutrinos have  nonzero masses---see e.g.~\cite{Xing:2020ijf} for a recent review. The effect of neutrino masses on the long-range forces becomes non-negligible when the distance $r$ exceeds $1/m_\nu$ where $m_{\nu}$ is the mass of the neutrino mediating the force. Including the mass effect, neutrino forces can be expressed in terms of the modified Bessel function~\cite{Grifols:1996fk}, which in the long-range limit ($r\gg 1/m_\nu^{}$)
decreases exponentially as ${\rm exp}\left(-2m_\nu^{}r\right)$. It is noteworthy that the effective potentials mediated by Dirac and Majorana neutrinos are different when neutrino masses cannot be neglected~\cite{Grifols:1996fk}, thus neutrino forces in principle could be used to determine the nature of neutrinos~\cite{Segarra:2020rah,Costantino:2020bei}. In addition to the mass effect, further generalizations including the flavor mixing have been addressed in Refs.~\cite{Lusignoli:2010gw,LeThien:2019lxh}. In Ref.~\cite{Asaka:2018qfg}, the contribution from the neutrino forces to the muonium hyperfine structure in the full SM was calculated.

% {\color{blue}[inserted from the intro.]
% The dispersion techniques developed by Feinberg and Sucher~\cite{Feinberg:1968zz,Feinberg1965,Feinberg:1989ps} (see also Ref.~\cite{Segarra:2015mqp} for a more pedagogical introduction) have been widely used in calculating the quantum forces mediated by exchange of different types of particles other than neutrinos, e.g., a pair of photons~\cite{Feinberg:1988yw,Feinberg:1991ac}, pseudo-scalars~\cite{Grifols:1994zz,Ferrer:1998ue}, etc.
% }

Going beyond the standard Fermi interactions, one might consider neutrino forces between other particles such as dark matter (DM)~\cite{Orlofsky:2021mmy}, neutrino forces modified by neutrino magnetic moments~\cite{Lusignoli:2010gw}, or, instead of neutrinos, other light particles 
% from a dark sector [XJ: removed to include]
mediating similar long-range forces~\cite{Fichet:2017bng,Brax:2017xho,Costantino:2019ixl,Banks:2020gpu, 
Feinberg:1988yw,Feinberg:1991ac,Grifols:1994zz,Ferrer:1998ue}. In Ref.~\cite{Orlofsky:2021mmy}, it has been shown that neutrino forces on DM could be strong enough to impact small-scale structure formation in the early 
universe\footnote{
	Cosmological effects of neutrino forces have been considered in early studies~\cite{Hartle:1970ug,Horowitz:1993kw,Ferrer:1998ju}. It is also worth mentioning that the many-body effect of neutrino forces in neutron stars once triggered a debate on the lower bound of $m_{\nu}$~\cite{Fischbach:1996qf,Smirnov:1996vj,Abada:1996nx,Kachelriess:1997cr,Kiers:1997ty,Abada:1998ti,Arafune:1998ft}.
	}. 
Refs.~\cite{Fichet:2017bng,Banks:2020gpu} systematically studied long-range forces arising from the exchange of two light particles of spin 0, 1/2, or 1. Ref.~\cite{Fichet:2017bng} showed that such forces could be used to search for DM in molecular spectroscopy and neutron scattering. 
It is interesting to note, as a conclusion of Ref.~\cite{Banks:2020gpu}, that these potentials are necessarily attractive if the contact interactions are of the scalar form, regardless of whether the light particles are fermions or bosons.

In the aforementioned works, only contact interactions are considered, which for massless neutrinos necessarily lead to the $1/r^5$ form. Since such forces have been used to compute non-relativistic scattering cross sections of DM or nucleons where the momentum transfer might be considerable (see e.g.~\cite{Orlofsky:2021mmy,Fichet:2017bng}), we take one step forward by asking to what extent the $1/r^5$ form remains valid at much smaller distances. 

The  $1/r^5$ form arises from the contact interaction, which can be seen by counting  dimensions in Eq.~\eqref{eq:SM nu-force}.  In the SM, as long as the external fermions are non-relativistic (otherwise the potential could not be well defined), the momentum transfer should be much smaller than the masses of $W$ and $Z$ bosons, which justifies the use of the contact interaction. As a consequence, when $r$ decreases, the $1/r^5$ form remains valid until $r$ approaches the inverse of the external fermion mass, in which case the non-relativistic approximation becomes invalid. 
% We also mention that the contribution from the neutrino forces to the muonium hyperfine structure in the full SM was calculated in Ref.~\cite{Asaka:2018qfg}.
%XJ: moved to ...

For non-SM interactions or particles (e.g.~DM), however, the contact interaction might be invalid before the non-relativistic approximation fails. In this case,  one needs to open the contact vertex between neutrinos and external fermions and recalculate the neutrino potential in a full renormalized theory. As we will show, if the contact vertex opens with a $t$-channel mediator $\phi$, in the short-range limit where the mass of $\phi$ is negligible compared with the momentum transfer, one may expect that the potential reduces to a Coulomb-like form varying as $1/r$. Therefore, it is tempting to formulate a unified framework that describes the variation of the potential from $1/r_{}^5$ (when $r\gg 1/m_\phi^{}$) to $1/r$ (when $r\ll 1/m_\phi^{}$), which will be presented in this work.

At the end of this introduction, we would like to briefly comment on the detection of neutrino forces. The SM neutrino force is extremely weak. From Eq.~(\ref{eq:SM nu-force}), one can estimate that only within the range of $10_{}^{-8}$ cm can it overcomes gravity. Nevertheless, new physics might enhance it to experimentally accessible levels. Recently it has been proposed that in atomic systems, the existing constraints on neutrino forces from experiments that search for new macroscopic forces will be significantly improved on and the future spectroscopy experiments will hopefully probe such forces~\cite{Stadnik:2017yge,Ghosh:2019dmi}.

The remainder of this paper is organized as follows. In Sec.~\ref{sec:contact interaction revisited}, we briefly revisit Feinberg and Sucher's formalism to calculate the long-range neutrino potential from contact interactions. In Sec.~\ref{sec:short-range behavior}, we embed the contact vertex into a full renormalized theory and investigate the short-range behavior of the neutrino forces. We consider two possible scenarios to open up the contact vertex by introducing a $t$-channel (Sec.~\ref{sub:t}) and an $s$-channel (Sec.~\ref{sub:s}) mediator. In both scenarios, we derive a general formula that is valid to describe the potential in all regimes as long as the external particles stay non-relativistic. We also show that the general formula will reduce to Feinberg and Sucher's $1/r_{}^5$ result in the long-range limit. Our main results and conclusions are summarized in Sec.~\ref{sec:summary}. Finally, technical details about on-shell renormalization and calculation of the Fourier integrals using the discontinuity of the amplitude are given in two appendices.

\section{Neutrino forces from contact interactions}
\label{sec:contact interaction revisited}
In this section, we briefly review Feinberg and Sucher's formalism~\cite{Feinberg:1968zz,Feinberg1965,Feinberg:1989ps} to calculate the neutrino potential from a contact interaction. 
For simplicity, we assume a scalar-type interaction between neutrinos and the external fermions $\chi$ with mass $m_\chi^{}$
\begin{eqnarray}
{\cal L}_{\rm int}^{}\supset G_{S}^{}\bar{\chi}\chi\bar{\nu}\nu\;,
\end{eqnarray}
where $G_{S}^{}$ is a dimensional coupling constant. Here for simplicity we consider only one generation of neutrino and the interaction is non-chiral. Generalizations to chiral interactions or interactions with other Lorentz structures are straightforward. In addition, since we mainly focus on the short-range behavior of the potential, it is safe to neglect the mass of neutrino.

According to the Born approximation, the effective potential between two $\chi$ particles (or a pair of $\chi$ and $\overline{\chi}$ particles)\footnote{For scalar interactions, the resulting potentials are independent of whether they are $\chi$ or $\overline{\chi}$ particles, and the potentials are always attractive. Vector interactions, by contrast, lead to repulsive potentials between two $\chi$ particles of the same charge. This is similar to the fact that tree-level scalar or vector mediators cause attractive or repulsive forces, see e.g.~\cite{Smirnov:2019cae}. } 
is related to the non-relativistic elastic scattering amplitude $\chi\chi\to\chi\chi$ (or $\chi\overline{\chi}\to\chi\overline{\chi}$) by the Fourier transform 
\begin{eqnarray}
	\label{eq:Born appro}
	V(\vec{r}\,)=-\int \frac{d^3 \vec{q}}{(2\pi)^3}e_{}^{i\vec{q}\cdot\vec{r}}{\cal A}_{\rm NR}^{}\left(\vec{q}\,\right),
\end{eqnarray}
where $\vec{q}$ is the momentum transfer and ${\cal A}_{\rm NR}^{}\equiv {\cal M}_{\rm NR}^{}/\left(4m_\chi^2\right)$ is the (normalized) scattering amplitude in the non-relativistic limit, with the factor $1/\left(4m_\chi^2\right)$ coming from normalization. For spin-independent scattering, 
${\cal A}_{\rm NR}^{}$ is only a function of $\rho\equiv \left|\vec{q}\,\right|$. Thus
one can integrate out the angular part first to obtain a central potential
\begin{eqnarray}
	\label{eq:fourier trans}
	V\left(r\right)=\frac{i}{4\pi^2 r}\int_{0}^{\infty}d\rho \rho {\cal A}_{\rm NR}^{}\left(\rho_{}^2\right)\left(e_{}^{i\rho r}-e_{}^{-i\rho r}\right)=\frac{i}{4\pi^2 r}\int_{-\infty}^{\infty}d\rho \rho {\cal A}_{\rm NR}^{}\left(\rho_{}^2\right)e_{}^{i\rho r}\;,
\end{eqnarray}
with $r\equiv \left|\vec{r}\right|$. 

\begin{figure}[t!]
	\centering
	\includegraphics[width=0.6\textwidth]{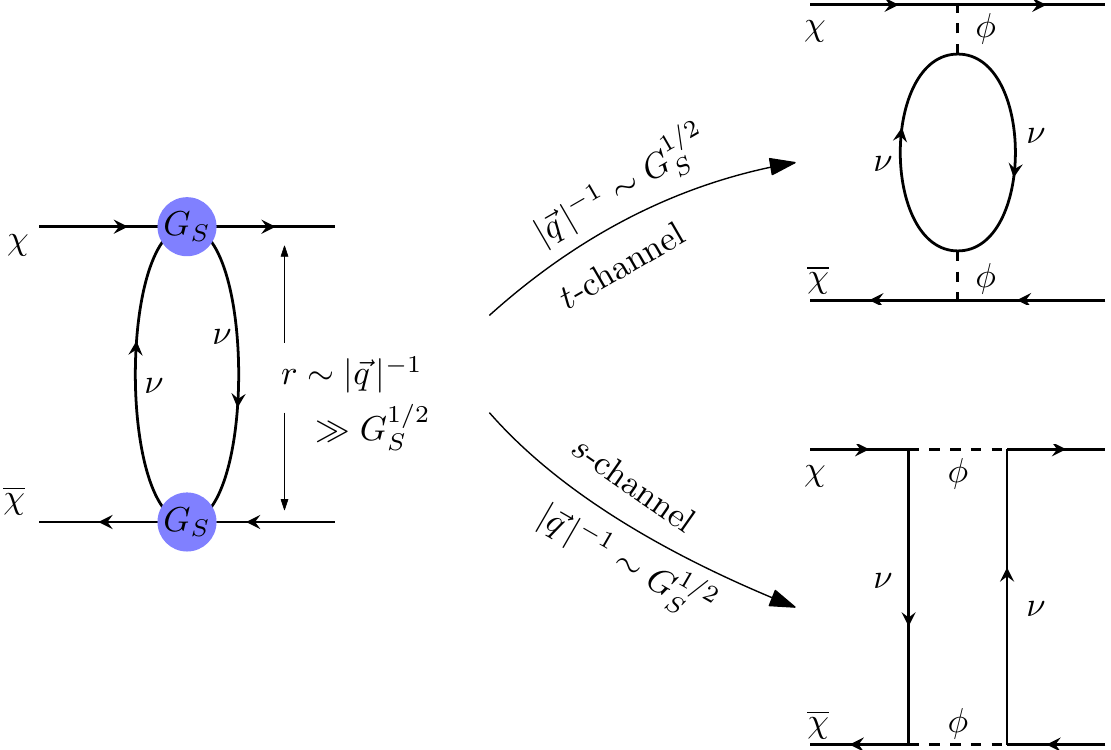}
	\caption{\label{fig:st} Neutrino forces from neutrino-pair exchange with a contact interaction (left diagram) and possible modifications (right diagrams) when the contact interaction is invalid due to $r\sim G_{S}^{1/2}$.} 
\end{figure}

The scattering amplitude can be derived by calculating the left loop diagram in  Fig.~\ref{fig:st}:
\begin{eqnarray}
i{\cal A}\left(q_{}^2\right)=-\left(iG_{S}^{}\right)_{}^2\int \frac{d^4k}{(2\pi)^4}{\rm Tr}\left(\frac{i}{\slashed{k}}\frac{i}{\slashed{k}+\slashed{q}}\right)=\frac{iG_{S}^2}{8\pi^2}q_{}^2\left[2+\Delta_{\rm E}^{}+\ln\left(\frac{\mu^2}{-q^2}\right)\right],
\end{eqnarray}
where $q$ is the transferred four-momentum, $\mu$ is the renormalization scale and $\Delta_{\rm E}^{}\equiv 1/\epsilon-\gamma_{\rm E}^{}+\ln\left(4\pi\right)$ with $\epsilon\to 0^{}_{+}$ and $\gamma_{\rm E}^{}\approx 0.577$ being the Euler–Mascheroni constant. Note that the normalized factor $1/(4m_\chi^2)$ has been cancelled by the non-relativistic wave functions of external fermions. In the non-relativistic limit, $q_{}^2\approx -\rho_{}^2<0$, thus
\begin{eqnarray}
\label{eq:contact amplitude}
{\cal A}_{\rm NR}^{}\left(\rho_{}^2\right)=-\frac{G_{S}^2}{8\pi^2}\rho_{}^2\left[2+\Delta_{\rm E}^{}-\ln\left(\frac{\rho^2}{\mu^2}\right)\right].
\end{eqnarray}
Substituting Eq.~(\ref{eq:contact amplitude}) into Eq.~(\ref{eq:fourier trans}), we have
\begin{eqnarray}
\label{eq:contact potential}
V\left(r\right)=-\frac{iG_{S}^2}{32\pi^4 r}\int_{-\infty}^{\infty}d\rho \rho_{}^3\left[2+\Delta_{\rm E}^{}-\ln\left(\frac{\rho^2}{\mu^2}\right)\right]e_{}^{i\rho r}=-\frac{3}{8\pi^3}\frac{G_{S}^2}{r^5}\;.
\end{eqnarray}
The Fourier integral can be computed by closing the contour on the complex plane of $\rho$\,---\,see Appendix~\ref{app:branch cut} for more details.

It is interesting to note that, for contact interactions, the integrand in Eq.~(\ref{eq:contact potential}) does not contribute to the final result except for the 
logarithmic term, which has a branch cut on the imaginary axis in the complex plane.
In particular, the UV divergent term does not appear in the effective potential, thus one only needs to extract the discontinuity at the branch cut (or equivalently, the imaginary part) of the amplitude and perform the integration along the imaginary axis.

\section{Short-range behavior of neutrino forces}
\label{sec:short-range behavior}
The neutrino-mediated potential derived from a contact interaction exhibits the $1/r^5$ behavior when $r$ is large. As $r$ decreases, the momentum transfer $\left|\vec{q}\right|\,\sim r^{-1}$ between the two external fermions increases and can eventually exceed the energy scale of $G_S^{-1/2}$, above which the contact interaction is invalid. Therefore, when $r$ is close to $G_S^{1/2}$ and the interaction becomes non-contact, we expect that the $1/r^5$ form should be altered.  

To investigate the short-range behavior of neutrino forces, we consider two possible scenarios for how the contact vertex may open up by introducing a $t$-channel (the upper right diagram in Fig.~\ref{fig:st}) or an $s$-channel (the lower right diagram) mediator. 

\subsection{The $t$-channel behavior}
\label{sub:t}
\begin{figure}[t!]
		\centering
		\includegraphics[width=0.98\textwidth]{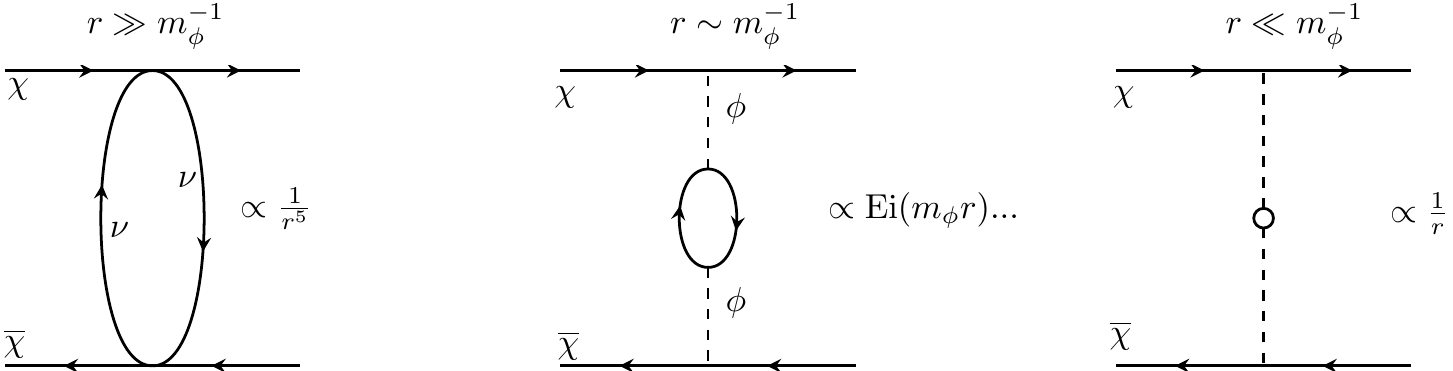}
		\caption{\label{fig:varied} A diagrammatic explanation of the $t$-channel neutrino potential varying from the $1/r^5$ to the $1/r$ forms. In the long-range limit ($r\gg m_\phi^{-1}$), the potential decreases as $1/r_{}^5$. At smaller distances ($r\sim m_{\phi}^{-1}$), the contact interaction vertices open up as two $\phi$ propagators. The resulting potential can be formulated in terms of the exponential integral function, ${\rm Ei}(m_\phi r)$---see Eqs.~(\ref{eq:loop potential})-(\ref{eq:loop potential normalized}). Further reducing $r$ to the short-range limit ($r\ll m_\phi^{-1}$),  the neutrino loop becomes subdominant and the potential reduces to the Yukawa/Coulomb-like form,  $1/r$.
		}
	%   \caption{A unified description of the behavior of the $t$-channel neutrino potential from $r\gg 1/m_\phi^{}$ to $r\ll 1/m_\phi^{}$ is established using the exponential integral function. The general potential formula is given in Eqs.~(\ref{eq:loop potential})-(\ref{eq:loop potential normalized}). In particular, in the long-range limit, the potential decreases with $1/r_{}^5$, while in the short-distance limit, the potential evolves with 
	% %   ${\rm ln}(m_\phi^{}r)/r$.} %[XJ: let's keep it simple]
	% $1/r$.{\color{red}[XJ's job: make the caption more explanatory.]}
	% }
\end{figure}

Let us first consider that the effective vertex is generated by a $t$-channel mediator. The Lagrangian reads:
\begin{eqnarray}
	{\cal L}_{\rm int}^{}\supset y_{\nu}\bar{\nu}\nu\phi+y_{\chi}\bar{\chi}\chi\phi\;,\label{eq:t-lagranian}
\end{eqnarray}
where $\phi$ is a real scalar of mass $m_\phi$, $y_{\nu}$ and $y_{\chi}$ denote the Yukawa couplings of $\phi$ to $\nu$ and $\chi$, respectively.

% Here for the purpose of illustration and simplicity we have assumed there is only one generation of neutrino and all the interactions are non-chiral and scalar-type. [XJ: moved to sec. 2]
% The generalization to the cases of chiral couplings and interactions with other Lorentz structures is straightforward. [XJ: moved to sec. 2]
% In addition, since we mainly focus on the small-distance behavior of the potential, it is safe to neglect the mass of neutrino. [XJ: moved to sec. 2]

In the long-range limit, $\phi$ can be integrated out and we obtain an effective four-fermion interaction, which reproduces Feinberg and Sucher's result: $V(r)\sim G_S^2/r^5$ with $G_S=y_{\nu} y_{\chi}/m_{\phi}^2$. This corresponds to the first diagram in Fig.~\ref{fig:varied}. Now consider that $r$ decreases to small values so that $|\vec{q}\, |$ is comparable to $m_{\phi}$. In this case, the contact interaction vertices are replaced by two $\phi$ 
% propagators [$i/(-m_{\phi}^2) \rightarrow i/(q^2-m_{\phi}^2)$, corresponding to the second diagram] and 
propagators, corresponding to the second diagram. And 
the potential exhibits complicated $r$ dependence involving the exponential integral function, ${\rm Ei}(m_\phi r)$. Further reducing $r$, we expect that the $\phi$ propagators will play a more important role than the neutrino loop---illustrated by the last diagram.  Eventually, with sufficiently small $r$, the effective potential is expected to be Yukawa- or Coulomb-like,  $V\sim 1/r$.

Below we will show, via explicit calculations,  that the potential indeed varies in this way.
% On the other hand, if $r\ll 1/m_{\chi_i}^{}$, then the external fermions are unable to maintain non-relatvistic and it is meaningless to define an effective potential between them. Therefore, we are mainly interested in the region where the external fermions keep non-relativistic while the scalar cannot be integrated out, i.e., $1/m_{\chi_i}^{} \ll r \lesssim 1/m_\phi^{}$.

% The effective potential is given by the Fourier transform of the non-relativistic scattering amplitude. 

% ,
First,  we need to compute the non-relativistic scattering amplitude. The neutrino loop in this case is also UV divergent, similar to the previous calculation in Sec.~\ref{sec:contact interaction revisited}. However, unlike the contact interaction case where the divergent and constant terms can be simply ignored, the UV divergence in this case requires a more careful treatment because, in addition to the branch cut from the logarithmic term, the scalar mediator contributes another type of singularities at $q^2=m^2$. As a consequence, constant terms arising from the loop also contribute. Therefore, we need to identify the physical part of the constant terms, which can be obtained by embedding the one-particle-irreducible (1PI) diagram into the physical propagator of $\phi$ and performing on-shell renormalization.
After the on-shell renormalization (see Appendix~\ref{app:renorm} for the details), we obtain the following scattering amplitude:
\begin{eqnarray}
\label{eq:amp finite}
i{\cal A}\left(q_{}^2\right)=\frac{iy_{\chi}^2y_{\nu}^2}{8\pi^2\left(q^2-m_\phi^2\right)^2}\left[
\left(q_{}^2-m_\phi^2\right)-q_{}^2{\rm ln}\left(\frac{-q^2}{m_\phi^2}\right)
\right].
\end{eqnarray}
In the non-relativistic limit, $q_{}^2\approx-\rho_{}^2<0$ with $\rho\equiv \left|\vec{q}\right|$, we have
\begin{eqnarray}
\label{eq:NR amp}
{\cal A}_{\rm NR}^{}\left(\rho_{}^2\right)=\frac{y_{\chi}^2y_{\nu}^2}{8\pi^2\left(\rho^2+m_\phi^2\right)^2}\left[-\left(\rho_{}^2+m_\phi^2\right)+\rho_{}^2{\rm ln}\left(\frac{\rho^2}{m_\phi^2}\right)\right],
\end{eqnarray}
which, after substituting into the Fourier transform (\ref{eq:fourier trans}), gives rise to the effective potential below (for the detailed calculation, see Appendix~\ref{app:branch cut}):
\begin{eqnarray}
\label{eq:loop potential}
V_{\rm loop}\left(r\right)&=&\frac{iy_{\chi}^2y_{\nu}^2}{32\pi^4r}\int_{-\infty}^{\infty}d\rho\frac{\rho}{\left(\rho^2+m_\phi^2\right)^2}\left[-\left(\rho_{}^2+m_\phi^2\right)+\rho_{}^2{\rm ln}\left(\frac{\rho^2}{m_\phi^2}\right)\right]e_{}^{i\rho r}\nonumber\\
	&=&\frac{m_\phi y_{\chi}^2 y_{\nu}^2}{64\pi^3}{\mathscr V}\left(m_\phi^{} r\right),
\end{eqnarray} 
where 
\begin{eqnarray}
\label{eq:loop potential normalized}
{\mathscr V}\left(x\right)\equiv \frac{2+e^x\left(2+x\right){\rm Ei}\left(-x\right)+e^{-x}\left(2-x\right){\rm Ei}\left(x\right)}{x}\;,
\end{eqnarray}
and ${\rm Ei}\left(x\right)\equiv-\int_{-x}^{\infty}dt \frac{e^{-t}}{t}$ is the exponential integral function. 
% Before diving into the analysis of the full potential, let us take a look at its asymptotic behaviors.
% \subsubsection{Asymptotic behaviors of the potential}

The large and small $x$ limits of ${\rm Ei}(x)$ are given as follows:
\begin{itemize}
	\item for $0<x\ll 1$, ${\rm Ei}(x)=\left(\gamma_{\rm E}+{\ln}x\right)+x+{\cal O}\left(x^2\right)$, ${\rm Ei}(-x)=\left(\gamma_{\rm E}+{\ln}x\right)-x+{\cal O}\left(x^2\right)$\;;
	\item for $x\gg 1$, ${\rm Ei}(x)=e^x\left[\frac{1}{x}+{\cal O}\left(\frac{1}{x^2}\right)\right]$, ${\rm Ei}(-x)=e^{-x}\left[-\frac{1}{x}+{\cal O}\left(\frac{1}{x^2}\right)\right]$\;.
\end{itemize}
With the above limit, it is straightforward to obtain the asymptotic behaviors of ${\mathscr V}\left(x\right)$:
\begin{eqnarray}
	{\mathscr V}\left(x\right)=
	\begin{cases}
		\frac{2}{x}\left(1+2\gamma_{{\rm E}}+2{\rm ln}x\right)+{\cal O}\left(x\right) & (\ensuremath{x\ll1})\\[2mm]
		-\frac{24}{x^{5}}+{\cal O}\left(\frac{1}{x^{7}}\right) & (\ensuremath{x\gg1})
	\end{cases}\,.
\end{eqnarray}

Therefore, in the long-range limit,  we have recovered Feinberg and Sucher's $1/r_{}^5$ potential
\begin{eqnarray}
\label{eq:long-range potential}
V_{\rm loop}\left(r\right)=-\frac{3 y_{\chi}^2y_{\nu}^2}{8\pi^3 m_\phi^4 r^5}\;,\quad r\gg m_\phi^{-1}\;,
\end{eqnarray}
which exactly matches the result of contact interaction in Eq.~(\ref{eq:contact potential}) with $G_{S}^{}=y_{\nu} y_{\chi} /m_\phi^2$.  
% In fact, this is equivalent to integrating out $\phi$ in the Lagrangian Eq.~(\ref{eq:lagranian}) from the beginning to give an effective operator $(y_{\chi} y_{\nu}/m_\phi^2)\bar{\chi}\chi\bar{\nu}\nu$. [XJ: moved to ...]

In the small-distance limit,  the potential evolves with ${\rm ln} (r)/r$:
\begin{eqnarray}
\label{eq:small-distance potential}
V_{\rm loop}\left(r\right)=-\frac{y_{\chi}^2y_{\nu}^2}{32\pi^3r}\left[2\,{\rm ln}\left(\frac{1}{m_\phi^{}r}\right)-1-2\gamma_{\rm E}^{}\right],\quad m_{\chi}^{-1}\ll r\ll m_\phi^{-1}\;.
\end{eqnarray}
One may observe that the potential in Eq.~\eqref{eq:small-distance potential} becomes divergent if $m_\phi^{}$ vanishes. 
This IR divergence can be removed when the neutrino mass effect is included. 

% The calculation of $t$-channel one-loop neutrino potential may remind one of the Uehling potential in the quantum electrodynamics, which modifies the Coulomb potential by the exchange of a pair of electrons. In fact, one can check that if the finite neutrino mass $m_\nu^{}$ is included, then the $t$-channel loop potential would also decrease as ${\rm exp}(-2m_\nu r)$ in the long-range limit $r\gg 1/m_\nu^{}$, similar to the case of Uehling potential. However, since we are mainly focusing on the short-range behavior of the neutrino forces in the present work, we can simply neglect the neutrino mass and keep $m_\phi^{}$ nonvanishing.
The calculation of the $t$-channel one-loop neutrino potential can be compared with 
the Uehling potential in QED. 
The Uehling potential is the radiative correction to the Coulomb potential. The vacuum polarization of the photon involves an electron loop and leads to a correction to the Coulomb potential proportional to $\exp(-2m_e r)$ when $r\gg 1/m_e$ with $m_e$ the electron mass.
This is similar to the neutrino loop in our calculation. 
In fact, one can check that if the finite neutrino mass $m_\nu^{}$ is included, then the 
potential here 
would also decrease as ${\rm exp}(-2m_\nu r)$ in the long-range limit when $r\gg 1/m_\nu^{}$, similar to the case of Uehling potential. However, since we are mainly focusing on the short-range behavior of the neutrino forces in the present work, we can simply neglect the neutrino mass and keep $m_\phi^{}$ nonvanishing.

\begin{figure}[t!]
	\centering
	\includegraphics[width=0.99\textwidth]{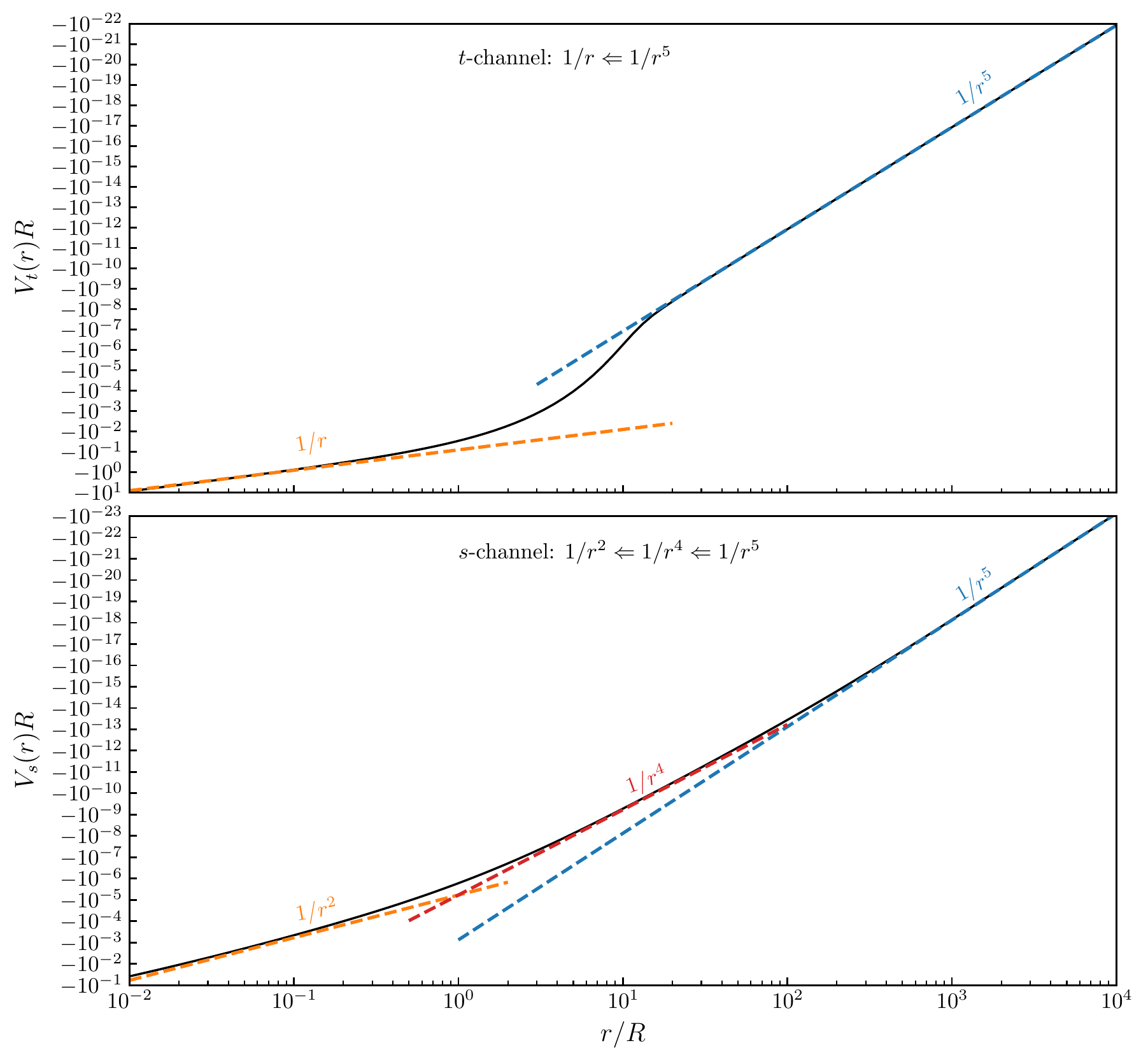}
	\caption{\label{fig:r5421} The evolution of the $t$-channel (upper panel) and $s$-channel (lower panel) effective potentials. In the long-range limit, both of the potentials decrease as $1/r^5$, while in the short-range limit, they exhibit  $1/r$, $1/r^2$, or $1/r^4$ behaviors, depending on the UV completion of the contact interactions and relevant mass scales---see the text for further details and discussions. The length unit $R$ is determined by $m_\phi$ and/or $m_\chi$ according to  Eqs.~\eqref{eq:Rt} and \eqref{eq:Rs}. }
	% $r\gg R$ where $R=1/m_{\phi}$ ($t$-channel) or $1/\sqrt{m_{\phi}^2-m_{\chi}^2}$ ($s$-channel),
\end{figure}

Note that in the presence of the above Yukawa interactions, $\phi$ can directly mediate a Yukawa potential between two $\chi$ particles:
\begin{eqnarray}
\label{eq:yukawa}
	V_{{\rm tree}}\left(r\right)=-\frac{y_{\chi}^{2}}{4\pi r}e^{-m_{\phi}r}\;.
\end{eqnarray} 
At $r\gg 1/m_{\phi}$, the tree-level force is exponentially suppressed and the loop-level long-range neutrino force dominates. At small distances, both contribute while the neutrino force is, compared to Eq.~\eqref{eq:yukawa},  weaker by a factor of $y_{\nu}^{2}/(16\pi^{2})$, which is the typical magnitude of loop suppression.
Therefore, our calculations of the neutrino force induced by the neutrino loop  can be alternatively viewed as a loop correction to the tree-level Yukawa force. 

In the spirit of a complete description of the potential, let us include the neutrino loop into the the propagator of $\phi$:
\begin{equation}
\frac{i}{q^{2}-m_{\phi}^{2}}\ \rightarrow\ \frac{i}{q^{2}-m_{\phi}^{2}+\Sigma(q^{2})}\;,\label{eq:propa}
\end{equation}
% <<<<<<< HEAD
where $\Sigma(q^{2})$ is the 1PI contribution. 
Computing the Fourier transform of the above loop-corrected propagator, one would obtain an effective potential that contains both contributions
from the tree-level Yukawa force and the loop-level neutrino force. They correspond to the leading order and next-to-leading order of
the expansion in $\Sigma(q^{2})$:
\begin{equation}
\frac{i}{q^{2}-m_{\phi}^{2}+\Sigma(q^{2})}\approx\frac{i}{q^{2}-m_{\phi}^{2}}-\frac{i}{\left(q^{2}-m_{\phi}^{2}\right)^{2}}\Sigma(q^{2})+\cdots\thinspace.\label{eq:expan}
\end{equation}
Since our calculations are carried out only at the one-loop level, it is consistent to take only the first two terms in the expansion and take $\Sigma(q_{}^2)\approx \Sigma_2^{}(q_{}^2)$, where $\Sigma_2^{}(q_{}^2)$ is the self-energy of $\phi$ to the order of ${\cal O}\left(y_\nu^2\right)$. This implies that we can obtain the full potential by simply adding Eq.~\eqref{eq:yukawa} to Eq.~\eqref{eq:loop potential}:
\begin{eqnarray}
\label{eq:t-channel potential}
	V_{t}\left(r\right)=V_{{\rm tree}}^ {}\left(r\right)+V_{\rm loop}\left(r\right)=\frac{m_{\phi}y_{\chi}^{2}}{4\pi}\left[-\frac{e^{-m_{\phi}r}}{m_{\phi}r}+\frac{y_{\nu}^{2}}{16\pi^{2}}{\mathscr{V}}\left(m_{\phi}^ {}r\right)\right].
\end{eqnarray}
Here $V_{t}(r)$ is the full potential, with the subscript $t$ indicating that it arises from the $t$-channel diagram. In the upper panel of Fig.~\ref{fig:r5421}, we present the full potential $V_{t}(r)$, together with its long- and short-range limits. In the shown example, we take $y_\nu=y_{\chi}=1$ and define the length scale $R$ as 
\begin{equation}
	R\equiv m_{\phi}^{-1}\ \  (\text{for \ensuremath{t}-channel})\;.\label{eq:Rt}
\end{equation}
It is clear that at short distances ($r\ll R$), the whole potential is dominated by the tree-level contribution, while in the long-range limit ($r\gg R$) it is dominated by the loop-level $1/r^5$ contribution since the tree-level Yukawa potential decreases exponentially with the distance.

\subsection{The $s$-channel behavior}
\label{sub:s}
Next, let us consider that the effective vertex opens with an $s$-channel
mediator---see the box diagram in Fig.~\ref{fig:st}. The Lagrangian
we consider is given by
\begin{eqnarray}
{\cal L}_{{\rm int}} \supset y\bar{\chi}\nu\phi+{\rm h.c.}\;,\label{eq:s-Lag}
\end{eqnarray}
where $\phi$ is a complex scalar field of mass $m_{\phi}$. We note
here that after integrating out $\phi$ and rearranging the fermionic
fields via Fierz transformations, Eq.~\eqref{eq:s-Lag} gives rise
to not only a contact interaction of the scalar form but also other
contact interactions such as vector and tensor interactions. Since
in this work we are interested in the $r$ dependence rather than
the spin dependence, the inclusion of these additional contact interactions
is beyond the scope of this work. 

With the interaction in Eq.~\eqref{eq:s-Lag} and the box diagram
in Fig.~\ref{fig:st}, we write down the amplitude of $\chi\bar{\chi}\to\chi\bar{\chi}$
scattering:
\begin{equation}
i{\cal M}=-\left|y\right|^{4}\int\frac{d^{4}k}{(2\pi)^{4}}\bar{v}\left(p_{2}\right)\frac{i}{\slashed{k}}u\left(p_{1}\right)\bar{u}\left(p_{3}\right)\frac{i}{\slashed{k}-\slashed{q}}v\left(p_{4}\right)\frac{i}{\left(p_{1}-k\right)^{2}-m_{\phi}^{2}}\frac{i}{\left(p_{2}+k\right)^{2}-m_{\phi}^{2}}\;,\label{eq:M1}
\end{equation}
where $p_{1}$ ($p_{3}$) and $p_{2}$ ($p_{4}$) denote, respectively,
the momenta of initial (final) momenta of $\chi$ and $\overline{\chi}$. In addition, the extra minus sign comes from the odd-number interchanges of anticommuting operators. Since the wave functions ($u$, $v$) only depend on the external
momenta, they can be extracted out of the integral, leading to
\begin{eqnarray}
i{\cal M}=-\left|y\right|_{}^{4}\bar{v}\left(p_{2}^ {}\right)\gamma_{\mu}u\left(p_{1}^ {}\right)\bar{u}\left(p_{3}^ {}\right)\gamma_{\nu}v\left(p_{4}\right)I^{\mu\nu}\;,\label{eq:M2}
\end{eqnarray}
where 
\begin{eqnarray}
I^{\mu\nu}=\int\frac{d^{4}k}{(2\pi)^{4}}\frac{k^{\mu}\left(k-q\right)^{\nu}}{k^{2}\left(k-q\right)^{2}\left[\left(p_{1}-k\right)^{2}-m_{\phi}^{2}\right]\left[\left(p_{2}+k\right)^{2}-m_{\phi}^{2}\right]}\;.\label{eq:tensor int}
\end{eqnarray}
Note that, as a common feature of box diagrams, the loop integral
is UV finite. 

After performing the loop integral in Eq.~(\ref{eq:tensor int}),
we obtain terms proportional to $g^{\mu\nu}$, $q^{\mu}q^{\nu}$,
$p_{i}^{\mu}p_{j}^{\nu}$, $p_{i}^{\mu}q^{\nu}$, $p_{i}^{\nu}q^{\mu}$, with
$i,j=1,2$. It can be shown that by taking the leading order of the non-relativistic
limit, only the $g^{\mu\nu}$ term contributes. For the $g^{\mu\nu}$
term, we take only the spin-independent\footnote{The contribution from spin-dependent terms to the neutrino forces can
be averaged out in spin summation.} part and obtain the following potential:
\begin{eqnarray}
V_{s}\left(r\right) & = & -\frac{3\left|y\right|^{4}}{128\pi^{3}r}\int_{0}^{\infty}dte_{}^{-\sqrt{t}r}\left[\frac{1}{2A}+\frac{B^{2}-tA}{4AB\sqrt{tA}}{\rm ln}\left(\frac{B-\sqrt{tA}}{B+\sqrt{tA}}\right)\right],\label{eq:V_s}
\end{eqnarray}
where $A\equiv t-4m_{\chi}^{2}$, $B\equiv t+2\Delta^{2}$ and
\begin{equation}
\Delta\equiv\sqrt{m_{\phi}^{2}-m_{\chi}^{2}}\;.\label{eq:delta}
\end{equation}
Here we have assumed $m_\phi^{}>m_\chi^{}$ otherwise $\chi$ would be unstable. Similar to Eq.~\eqref{eq:Rt}, we also define a length scale:
\begin{equation}
R\equiv\Delta^{-1}\ \ (\text{for \ensuremath{s}-channel})\;.\label{eq:Rs}
\end{equation}

The potential can be expanded in different ways to obtain different limits as we will present  below.
One should note, however, that the potential is only valid
in the range of $r\gg m_{\chi}^{-1}$ since the external fermions
should keep non-relativistic. Therefore, we may encounter several
possible hierarchies depending the comparison of $R$ with $m_{\chi}^{-1}$
and $r$. 

The simplest case is $R<m_{\chi}^{-1}$. In this case, we only need
to consider one possible hierarchy: $R<m_{\chi}^{-1}\ll r$. And the
result simply reduces to
\begin{eqnarray}
V_{s}\left(r\right)=-\frac{3\left|y\right|^{4}R^{4}}{128\pi^{3}r^{5}}\qquad\left(R<m_{\chi}^{-1}\ll r\right).\label{eq:V1}
\end{eqnarray}
It implies that the $1/r^{5}$ behavior remains valid as $r$ decreases
until $r$ approaches $m_{\chi}^{-1}$. 

If $m_{\chi}^{-1}<R$ (for better illustration, we consider $m_{\chi}^{-1}\ll R$,
which is the case in Ref.~\cite{Orlofsky:2021mmy}), then it becomes more complicated.
As $r$ decreases from sufficiently large values, the $1/r^{5}$ behavior
becomes invalid when $r$ approaches a point where $m_{\chi}^{-1}/R\sim R/r$
(i.e. $r\sim m_{\chi}R^{2}$). After passing this point, the $1/r^{5}$
behavior is altered to $1/r^{4}$. Further, when it passes $R$ (which
is smaller than $m_{\chi}R^{2}$), it changes to $1/r^{2}$ and remains
in this form until the non-relativistic approximation becomes invalid.
The short-, intermediate- and long-range limits of the potential are
given as follows:
\begin{equation}
V_{s}\left(r\right)=-\frac{3\left|y\right|^{4}}{128\pi^{3}}\times
\begin{cases}
\frac{\pi}{4m_{\chi}r^{2}} & \left(m_{\chi}^{-1}\ll r\ll R\right)\\[2mm]
\frac{\pi R^{2}}{4m_{\chi}r^{4}} & \left(R\ll r\ll m_{\chi}R^{2}\right)\\[2mm]
\frac{R^4}{r^{5}} & \left(m_{\chi}R^{2}\ll r\right)
\end{cases}\;.\label{eq:V_s-1}
\end{equation}

In the lower panel of Fig.~\ref{fig:r5421}, we present the full potential $V_{s}(r)$ given by Eq.~\eqref{eq:V_s}, together with the three limits in Eq.~\eqref{eq:V_s-1}. In the shown example, we set $y=1$ and $m_\chi=100/R$. Hence the valid ranges for the aforementioned short, intermediate- and long-range limits are $10^{-2} \ll r/R\ll 1$,  $1 \ll r/R\ll 10^{2}$,  and $10^{2} \ll r/R$, respectively.

\section{Summary}
\label{sec:summary}
In this paper, we investigated the short-range behavior of the neutrino forces which arise from the exchange of a pair of neutrinos between two fermions. Although in the SM, the interaction between neutrinos and external fermions can be effectively described by a contact vertex which leads to a $1/r_{}^5$ potential, this may be not the case in new physics scenarios (in particular, for DM interactions), where the contact interaction might be invalid before the non-relativistic approximation of external fermions fails. 
Thus it is necessary to study possible variations of the potential in such cases.
% Thus it is necessary to 
% embed the contact vertex into a renormalized theory and calculate the full potential.

We considered two possible scenarios to open up the contact vertex by introducing a $t$-channel or an $s$-channel mediator. In the $t$-channel scenario, the potential induced by the neutrino loop is given in Eqs.~(\ref{eq:loop potential})-(\ref{eq:loop potential normalized}). In the long-range limit, it decreases with $1/r_{}^5$ as expected while in the short-range limit it evolves as ${\rm ln}(m_\phi^{}r)/r$. Including the tree-level Yukawa contribution, the full potential in the $t$-channel scenario is given in Eq.~(\ref{eq:t-channel potential}) and its behavior is shown in the upper panel of Fig.~\ref{fig:r5421}. For $1/m_\chi^{}\ll r\ll 1/m_\phi^{}$, the full potential is dominated by the tree-level contribution which behaves as $1/r$, while for $r\gg 1/m_\phi^{}$ it is dominated by the loop-level $1/r_{}^5$ contribution since the tree-level Yukawa potential decreases exponentially.

In the $s$-channel scenario, the neutrino potential is induced by the box diagram, with the complete expression given in Eq.~(\ref{eq:V_s}) and an example shown in the lower panel of Fig.~\ref{fig:r5421}. The potential exhibits different behaviors as the distance approaches different limits: for the long-range limit $r\gg m_\chi^{}/(m_\phi^2-m_\chi^2)$, it behaves as $1/r_{}^5$ as expected; for the intermediated-range limit $1/\sqrt{m_\phi^2-m_\chi^2}\ll r\ll m_\chi^{}/(m_\phi^2-m_\chi^2)$, it behaves as $1/r_{}^4$; and for the short-range limit $1/m_\chi^{}\ll r\ll 1/\sqrt{m_\phi^2-m_\chi^2}$, it behaves as $1/r_{}^2$.

We emphasize that although the analysis in the present paper is performed only on the scalar-type and non-chiral interactions for the purpose of illustration and simplicity, the generalizations to the cases of chiral interactions or interactions with other Lorentz structures are straightforward. Our results might be of potential importance to the study of long-range force effects in particle physics and cosmology. 

We have neglected the neutrino mass effect throughout this paper since we mainly focus on the short-range behavior of neutrino forces. However, it will be very interesting 
to extend our results to the nonvanishing neutrino mass case and investigate the complete behavior of the potential in the full range, 
% We have neglected the neutrino mass effect throughout this paper since we mainly focus on the short-range behavior of neutrino forces. However, it will be of great interest to extend our results to the nonvanishing neutrino mass case and investigate the complete behavior of the potential in a full renormalized theory, 
especially in the non-relativistic neutrino environment like cosmic neutrino background, where the neutrino mass cannot be neglected and the temperature effects must be taken into account. We leave these interesting extensions for future works.

% We hope that our efforts in this work will play a role in the complete study of neutrino forces in particle physics beyond the SM.

\section*{Acknowledgements}
The authors would like to thank Rupert Coy, Jichen Pan, Michel Tytgat, Di Zhang, and Shun Zhou for helpful discussions. This work was supported in part by the National Natural Science Foundation of China under grant No. 11775232 and No. 11835013, by the Key Research Program of the Chinese Academy of Sciences under grant No. XDPB15, and by the CAS Center for Excellence in Particle Physics.

\begin{appendix}

\section{Renormalization of the $\phi$ propagator}
\label{app:renorm}
% In order to eliminate the infiniteness in the loop calculation, one should identify the scalar field and scalar mass in Eq.~(\ref{eq:t-lagranian}) as bare quantities $\phi_0^{}$ and $m_0^{}$, and they are related to the physical quantities $\phi$ and $m_\phi^{}$ by the counterterms~
The neutrino loop that contributes to the two-point function of $\phi$  in Sec.~\ref{sub:t} is UV divergent. 
In order to obtain its physical contribution, we shall renormalize the $\phi$ field and its mass.\footnote{The renormalization of $y_{\nu}$ does not affect the calculation of the self-energy of $\phi$, thus we can simply identify $y_{\nu}$ as the finite renormalized quantity in our calculation.} Following the standard renormalization procedure, we denote the bare field strength and the mass by $\phi_0^{}$ and $m_0^{}$, and the physical ones by $\phi$ and $m_\phi^{}$, respectively. The renormalization is hence formulated as
\begin{eqnarray}
\label{eq:re}
\phi_0^{}=Z_\phi^{}\phi\equiv\left(1+\delta_\phi^{}\right)\phi\;,\quad
m_0^{2}=Z_m^{}m_\phi^{2}\equiv\left(1+\delta_m^{}\right)m_\phi^{2}\;.
\end{eqnarray}
The bare self-energy of the scalar under the dimensional regularization, 
% to the accuracy of 
to the order of 
${\cal O}\left(y_{\nu}^2\right)$, is given by
\begin{eqnarray}
i\Sigma_2^0\left(q_{}^2\right)=-\left(iy_{\nu}\right)_{}^2\int \frac{d^4k}{\left(2\pi\right)^4}{\rm Tr}\left(\frac{i}{\slashed{k}}\frac{i}{\slashed{k}+\slashed{q}}\right)=\frac{iy_{\nu}^2}{8\pi^2}q_{}^2\left[2+\Delta_{\rm E}^{}+\ln\left(\frac{\mu^2}{-q^2}\right)\right],
\end{eqnarray}
where the extra minus sign comes from the neutrino loop, $\mu$ is the renormalization scale, $\Delta_{\rm E}^{}\equiv 1/\epsilon-\gamma_{\rm E}^{}+\ln\left(4\pi\right)$ with $\epsilon\to 0_+^{}$ and $\gamma_{\rm E}^{}\approx 0.577$ being the Euler–Mascheroni constant.

The renormalized self-energy is a combination of the bare one plus the counterterms arising from Eq.~\eqref{eq:re}:
\begin{eqnarray}
\label{eq:relation}
\Sigma_2^{}\left(q_{}^2\right)=\Sigma_2^0\left(q_{}^2\right)+q_{}^2 \delta_\phi^{}-m_{\phi}^2\left(\delta_\phi^{}+\delta_m^{}\right).
\end{eqnarray}
% The determination of counterterms depends on the subtraction scheme. Here we take the on-shell subtraction scheme, whose renormalization conditions are given by
The counterterms are determined by the on-shell renormalization conditions\footnote{Note that for unstable particles, since their on-shell self-energy contains imaginary parts, the on-shell renormalization conditions fix the real parts~\cite{Aoki:1982ed,Bohm:1986rj}.}:
\begin{eqnarray}
\label{eq:renorm condi}
{\rm Re}\,\Sigma_2^{}\left(q_{}^2\right)|_{q^2=m_\phi^2}^{}=0\;,\quad
\frac{d}{dq^2}{\rm Re}\,\Sigma_2^{}\left(q_{}^2\right)|_{q^2=m_\phi^2}^{}=0\;.
\end{eqnarray}
Substituting Eq.~(\ref{eq:relation}) into Eq.~(\ref{eq:renorm condi}), we solve it with respect to $\delta_m$ and $\delta_\phi$:
\begin{eqnarray}
\label{eq:counterterms}
\delta_m^{}&=&\frac{1}{m_\phi^2}{\rm Re}\,\Sigma_2^0\left(q_{}^2\right)|_{q^2=m_\phi^2}^{}=\frac{y_{\nu}^2}{8\pi^2}\left[2+\Delta_{\rm E}^{}+{\rm ln}\left(\frac{\mu^2}{m_\phi^2}\right)\right],\nonumber\\
\delta_\phi^{}&=&-\frac{d}{dq^2}{\rm Re}\,\Sigma_2^{0}\left(q_{}^2\right)|_{q^2=m_\phi^2}=-\frac{y_{\nu}^2}{8\pi^2}\left[1+\Delta_{\rm E}^{}+{\rm ln}\left(\frac{\mu^2}{m_\phi^2}\right)\right].
\end{eqnarray}
Then the renormalized self-energy is determined by substituting Eq.~(\ref{eq:counterterms}) back into Eq.~(\ref{eq:relation}):
\begin{eqnarray}
\label{eq:renorm selfenergy}
\Sigma_2^{}\left(q_{}^2\right)=\frac{y_{\nu}^2}{8\pi^2}\left[\left(q_{}^2-m_\phi^2\right)+q_{}^2{\rm ln}\left(\frac{m_\phi^2}{-q^2}\right)\right].
\end{eqnarray}
Note that the self-energy of $\phi$ in Eq.~(\ref{eq:renorm selfenergy}) has an imaginary part when it goes on-shell, $q_{}^2=m_\phi^2$. This corresponds to the instability of $\phi$: it can decay into a pair of neutrinos with the decay rate $\Gamma_\phi^{}=y_{\nu}^2m_\phi^{}/(8\pi)$. However, for the non-relativistic $t$-channel scattering process we consider throughout this paper, the momentum transfer $q$ is always spacelike, namely $q_{}^2<0$, thus the amplitude is always real. Finally, the renormalized amplitude is given by
\begin{eqnarray}
\label{eq:renorm amp}
i{\cal A}\left(q_{}^2\right)&=&\left(iy_{\chi}\right)^2\frac{i}{q^2-m_\phi^2}i\Sigma_2^{}\left(q_{}^2\right)\frac{i}{q^2-m_\phi^2}\nonumber\\
&=&\frac{iy_{\chi}^2 y_{\nu}^2}{8\pi^2\left(q^2-m_\phi^2\right)^2}\left[\left(q_{}^2-m_\phi^2\right)-q_{}^2{\rm ln}\left(\frac{-q^2}{m_\phi^2}\right)\right],
\end{eqnarray}
where the normalized factor $1/\left(4m_{\chi}^2\right)$ has been cancelled by the non-relativistic wave functions of external fermions. The amplitude in Eq.~(\ref{eq:renorm amp}) is finite.

\section{Branch cuts and singularities of the amplitude}
\label{app:branch cut}

In this appendix, we present in detail the calculations of the Fourier
integrals used in this work. 

For neutrino forces from contact interactions, we encounter the following integral:
\begin{align}
I_{c} & \equiv-\int\frac{d^{3}\vec{q}}{\left(2\pi\right)^{3}}e^{i\vec{q}\cdot\vec{r}-|\vec{q}\cdot\vec{r}|0_{+}}\left[C-q^{2}\ln\left(-\frac{q^{2}}{\mu^{2}}\right)\right]\nonumber \\
 & =\frac{i}{4\pi^{2}r}\int_{-\infty}^{\infty}d\rho e^{i\rho r}\rho\left[C+\rho^{2}\ln\left(\frac{\rho^{2}}{\mu^{2}}\right)\right]\nonumber \\
 & =\frac{i}{4\pi^{2}r}\int_{0}^{\infty}id\rho_{i}e^{-\rho_{i}r}i^3\rho_{i}^{3}\left[\ln\left(\frac{-\rho_{i}^{2}+i0_{+}}{\mu^{2}}\right)-\ln\left(\frac{-\rho_{i}^{2}-i0_{+}}{\mu^{2}}\right)\right]\nonumber \\
 & =\frac{i}{4\pi^{2}r}\int_{0}^{\infty}d\rho_{i}e^{-\rho_{i}r}\rho_{i}^{3}\times 2\pi i\nonumber \\
 & =-\frac{3}{\pi r^{5}}\thinspace.\label{eq:I-c}
\end{align}
Here $C$ is some quantity that does not have singularities or branch cuts on the imaginary axis of $\rho$, $q^{2}=q_{\mu}q^{\mu}=-|\vec{q}|^2$,  $\rho\equiv|\vec{q}|$, $\rho_i$ is the imaginary part of $\rho$ (on the imaginary axis, $\rho=i \rho_i$),  and $0_{+}$ denotes a positive infinitesimal number. The logarithmic part contains a branch cut on the imaginary axis. Hence in the third step we push the contour upward so that it becomes an integral along the imaginary axis.

\begin{figure}
\centering
\includegraphics[width=0.7\textwidth]{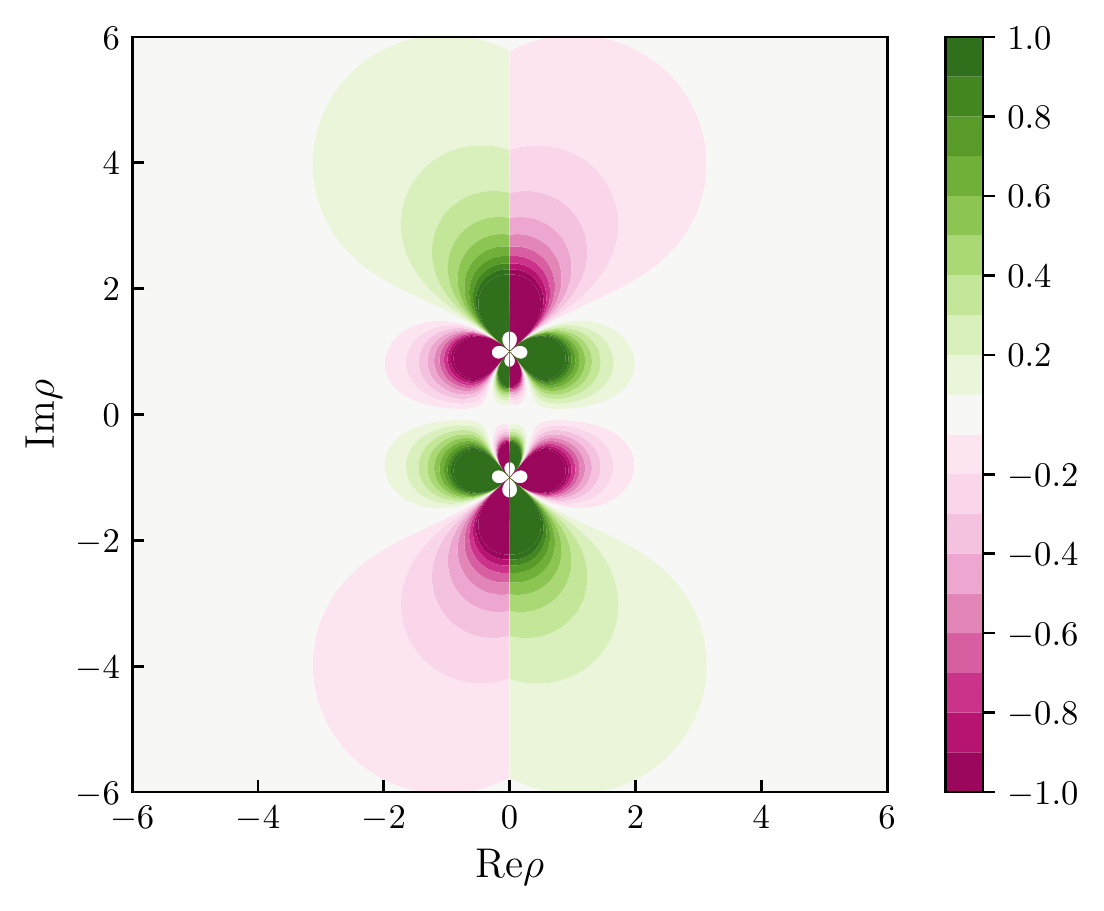}
\caption{ Contours of ${\rm Im}\left\{ \left(\rho^{2}+m^{2}\right)^{-2}\left[\rho^{2}\ln\left(\rho^{2}/m^{2}\right)-\rho^{2}-m^{2}\right]\right\}$, to show branch cuts (on the imaginary axis) and singularities (at $\rho = \pm i m$ with $m=1$) of the integrand in Eq.~\eqref{eq:I-t}.
\label{fig:cut}
}
\end{figure}

% $t$-channel $\phi$-$\nu\overline{\nu}$-$\phi$:
To study neutrino forces with the $t$-channel behavior, we need to compute the following integral:
\begin{align}
I_{t} & \equiv-\int\frac{d^{3}\vec{q}}{\left(2\pi\right)^{3}}\frac{e^{i\vec{q}\cdot\vec{r}-|\vec{q}\cdot\vec{r}|0_{+}}}{\left(q^{2}-m^{2}\right)^{2}}\left[q^{2}-m^{2}-q^{2}\ln\left(-\frac{q^{2}}{m^{2}}\right)\right]\nonumber \\
 & =\frac{i}{4\pi^{2}r}\int_{-\infty}^{\infty}d\rho\,e^{i\rho r}\rho\,\frac{\rho^{2}\ln\left(\frac{\rho^{2}}{m^{2}}\right)-\rho^{2}-m^{2}}{\left(\rho^{2}+m^{2}\right)^{2}}\thinspace.\label{eq:I-t}
\end{align}
The integrand contains both singularities (at $\rho=\pm im$) and
a branch cut on the imaginary axis, as shown in Fig.~\ref{fig:cut}.
To compute the integral, we slightly pull the singularities off the
imaginary axis so that the integral can be converted to an integral
around one of the singularities and an integral along the branch cut.
Specifically, we replace the integrand with the following function:
\begin{equation}
f(\rho,\ \epsilon)\equiv e^{i\rho r}\rho\,\frac{\rho^{2}\ln\left(\frac{\rho^{2}}{m^{2}}\right)-\rho^{2}-m^{2}}{\left(\rho^{2}+m^{2}-i\epsilon\right)^{2}},\ (\epsilon\to 0_+)\thinspace.\label{eq:I}
\end{equation}
The residue of $f$ at the $\rho=i\sqrt{m^{2}-i\epsilon}$ singularity
contributes the following part to the integral:\footnote{If we had pulled the singularities to the opposite direction by replacing
$i\epsilon$ with $-i\epsilon$ in Eq.~\eqref{eq:I}, the residue
in Eq.~\eqref{eq:I-1} would differ by a minus sign. However, as one
can check, the contribution of the branch cut in this case would also
change while the total contribution remains the same.}
\begin{equation}
2\pi i\lim_{\epsilon\rightarrow0_{+}}{\rm Res}\left[f(\rho=i\sqrt{m^{2}-i\epsilon},\ \epsilon)\right]=\frac{\pi^{2}}{2}e^{-mr}(mr-2)\thinspace.\label{eq:I-1}
\end{equation}
The branch cut along the positive part of the imaginary axis contributes
to the integral as follows:
\begin{align}
\int_{0}^{\infty}id\rho_{i}\lim_{\rho_{r}\rightarrow0_{+}}\left[f(i\rho_{i}+\rho_{r},\ \epsilon)-f(i\rho_{i}-\rho_{r},\ \epsilon)\right] & =2\pi i \int_{0}^{\infty}d\rho_{i}\frac{\rho_{i}^{3}e^{-\rho_{i}r}}{\left(\rho_{i}^{2}-m^{2}+i\epsilon\right)^{2}}\nonumber \\
 & =-\frac{\pi^{2}}{2}e^{-mr}(mr-2)-\frac{i\pi}{2}mr\mathscr{V}(mr)\thinspace,\label{eq:I-3}
\end{align}
where the $\mathscr{V}$ function was defined in Eq.~\eqref{eq:loop potential normalized}.

Combining the results in Eqs.~\eqref{eq:I-1} and \eqref{eq:I-3},
we obtain 
\begin{equation}
I_{t}=\frac{m}{8\pi}\mathscr{V}(mr)\thinspace.\label{eq:I-4}
\end{equation}

For the $s$-channel case, the corresponding Fourier integral cannot be expressed in terms of known special functions. However, we can follow similar steps to convert it to an integral along the imaginary axis. This leads to the integral in Eq.~\eqref{eq:V_s}.

\end{appendix}

\bibliographystyle{JHEP}
\bibliography{ref}
\end{document}